# Cylindrical vector beams for rapid polarization-dependent measurements in atomic systems

**F. K. Fatemi**

*Optical Sciences Division, Naval Research Laboratory, Washington DC, 20375*
*[coldatoms@nrl.navy.mil](mailto:coldatoms@nrl.navy.mil)*

**Abstract:** We demonstrate the use of cylindrical vector beams – beams with spatially varying polarization – for detecting and preparing the spin of a warm rubidium vapor in a spatially dependent manner. We show that a modified probe vector beam can serve as an atomic spin analyzer for an optically pumped medium, which spatially modulates absorption of the beam. We also demonstrate space-variant atomic spin by optical pumping with the vector beams. The beams are thus beneficial for making single-shot polarization-dependent measurements, as well as for providing a means of preparing samples with position-dependent spin.

## 1. Introduction

In recent years, significant attention has been devoted to the optical properties of cylindrical vector beams (CVBs), which are cylindrically-symmetric hollow beams that have radial or azimuthal polarization [1]. For example, a tightly focused radially polarized beam has a strong longitudinal field component that is smaller than achievable with linearly polarized light [2, 3]. The optical propagation and focusing characteristics have been examined in detail [2-6]. Their unique propagation characteristics and potential applications have led to several simple and inexpensive techniques for generating them [1, 2, 7-10] and they have been proposed for applications such as optical trapping [11], atom guiding [12], laser machining [13], charged particle acceleration [14-15], and polarimetry [16].

Yet despite numerous reports discussing their characteristics, generation, and potential applications, it has only been in very recent years that some of these ideas have been implemented and demonstrated experimentally. They have been used for probing molecules [3], trapping microspheres [11], generating surface plasmon resonances [17], entangling spatial and polarization degrees of freedom [18], and optical microscopy [19].

In this paper, we extend the domain of CVB interactions to polarization sensitive diagnostics and manipulation of atomic systems. Although much of the prior interest in these beams derives from their unique focusing behaviors, their use here is motivated by the spatially varying polarization profile. While earlier works have generally used spatially-varying *linear* polarization (radial or azimuthal), we modify the beams to have azimuthally-varying *elliptical* polarization so that the space-variant spin in the laser beam can be transferred to the atomic medium. Therefore, these modified CVBs offer a unique way to manipulate atoms spatially. Atomic systems can be probed with light carrying all degrees of elliptical polarization simultaneously, or prepared in a position-dependent way. Here, we demonstrate polarization multiplexing in warm rubidium vapor by preparing and detecting properties of the medium with CVBs. This could be of interest in coherent or nonlinear processes in two dimensions. For example, spatial modulation of light in atomic vapor has been proposed using input beams with spatially-varying intensity [20], and entanglement of multimode images using linearly polarized light has been shown in rubidium vapor [21].

## 2. CVB formation and beam quality

In the paraxial approximation, a CVB has either radial or azimuthal polarization:

$$\vec{E}_r(r) = E_0 r \exp\left(-\frac{r^2}{\omega_1^2}\right)\hat{r} \qquad (1)$$

$$\vec{E}_\varphi(r) = E_0 r \exp\left(-\frac{r^2}{\omega_1^2}\right)\hat{\varphi} \qquad (2)$$

where $\omega_1$ is the waist parameter and $\varphi$ is the azimuthal coordinate. We have generally found that fiber-based CVB generation [8, 9] is both more compact and stable compared to the SLMs used in our prior work [12, 22], with the added benefit that time-dependent fluctuations due to SLM refresh rate are eliminated. Though the technique here is similar to that of Ref. [9], we summarize it and the resulting beam qualities for our wavelength and fiber as we have found that these parameters markedly affect the output mode profiles.

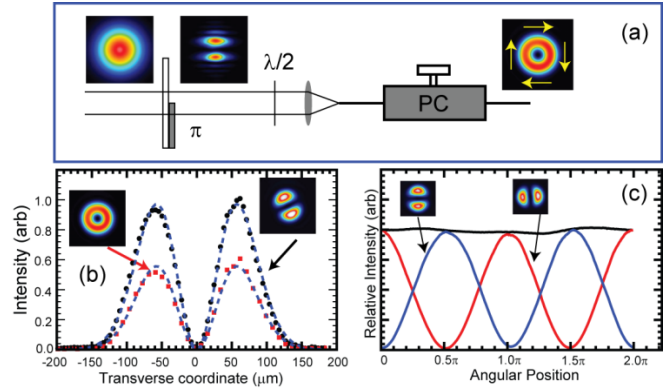

Fig.1. (a) Setup for CVB formation. A Gaussian beam passes through a π phase plate, producing a close approximation to a TEM$_{01}$ beam. A polarization controller (PC) and λ/2 waveplate adjust the profile in 1060 nm fiber. An output profile with azimuthal polarization is shown. (b) Two different output profiles formed by rotating the half-wave plate. The total power throughput is approximately the same (~10% more for the CVB than for the two-lobed profile). Fits to Eq. 1 are shown (dotted lines). (c) Profiles after passing the azimuthally-polarized CVB through a linear polarizer with horizontal (blue) and vertical orientation (red). Minima are ~1-2% of the maxima. Azimuthal intensity without any polarizers shown in black.

Fig. 1(a) shows the setup. A Gaussian beam of wavelength 795 nm passes through a π-phase plate formed by depositing 390 nm of SiN onto one half of a quartz plate. This approximately produces a TEM$_{01}$ free space mode, and is coupled into Corning HI 1060 optical fiber with specified cutoff wavelength $\lambda_c \approx 920$ nm and numerical aperture of 0.14. At 795 nm, this fiber supports only the fundamental HE$_{11}$ mode and first excited TM$_{01}$, TE$_{01}$, and HE$_{21}$ modes. Because of the π phase shift, there is strictly no overlap with the HE$_{11}$ mode for a well-aligned input beam so the light is in a superposition of the three other supported modes.

Mode conversion between these higher orders is effected with the half-wave plate in front of the fiber and the stress-induced birefringence of the polarization controller (PC). In principle, the waveplate is not needed (and was not used in Ref. [9]), but it provides a reproducible handle for control – the PC applies stress and torsion to the fiber, but there is noticeable hysteresis. The typical output from our mode converter is shown in Fig. 1(a). We also show, in Fig. 1(b), a two-lobed output beam for a different orientation of the waveplate, along with the CVB. The total power output does not change appreciably when varying the waveplate, yet the output profile can be adjusted from circular to lobed, indicating that mode conversion within the fiber is essentially lossless. At the fiber output, transverse cuts of the intensity fit well to $|E_r(r)|^2$ in Eq. 1. We note that although these higher order fiber modes in the core are Bessel functions and depend on the fiber geometry and refractive indices [23], our fiber and wavelength combination produces a mode that overlaps well with $|E_r(r)|^2$.

Because the conversion between the two-lobed profile and CVB profile occurs without loss of power, we can calculate an approximate overlap integral using the modified Gaussian beam and the two-lobed, $TEM_{01}$-like output in Fig. 1(b), which has a field profile $|E_r(r)|\sin(\varphi)\hat{y}$. Modifying the Gaussian with a $\pi$ phase plate yields:

$$\vec{E}_{MG} = E_1 \exp\left(-\frac{r^2}{\omega_0^2}\right) sign(y)\hat{y} \qquad (3)$$

where $y = r\sin(\varphi)$ and $sign(x) = x/|x|$. The optimum overlap integral occurs when $\omega_0 = 2^{1/2}\omega_1$, at which value the overlap integral is $\pi^{-1/2}(4/3)^{3/2} = 0.87$, or a power conversion of 75%. The exact overlap integral will depend on the specific fiber, due to deviations of the waveguide mode from that of Eq. 1, but for our fiber and wavelength this is a good approximation. With fiber coupling losses, we typically achieve over 50% conversion efficiency.

The specific fiber used is important for optimizing mode purity. If the fiber can accommodate more than just the first excited modes, mode selection is difficult. This is because the number of available modes grows with the fiber core area, so slight perturbations decompose the input beam into a superposition of a large number of modes. From $\lambda_c$, we can estimate the CVB cutoff wavelength $\lambda_{CVB} = 2.4/3.8\,\lambda_c$ [24], at which higher orders become available. For our fiber, with $\lambda_c \approx 920$ nm, $\lambda_{CVB} \approx 580$ nm, giving ~340 nm range of CVB operation before other modes can be excited. Mode propagation constants near cutoff conditions, however, are sensitive to small perturbations of the fiber geometry. We tested this using SMF-28 optical fiber, which has $\lambda_c \approx 1260$ nm, $\lambda_{CVB} = $ ~790 nm. With 795 nm input light, 0.6% away from $\lambda_{CVB}$, the stability of the CVB beam was severely compromised and it was more difficult to mode-select the CVB. At the other extreme, if $\lambda_c$ is very close to the operating wavelength, output intensity profiles can also deviate substantially from those presented here and have poor coupling efficiency. Finally, because instability grows with fiber length, we used short fiber sections. With our setup, a 15-cm-long Corning HI 1060 fiber operating at 795 nm, the output mode remains stable for several weeks.

Within the paraxial approximation, a CVB can be decomposed into the sum of orthogonally-polarized $TEM_{10}$ and $TEM_{01}$ modes. The purity of the CVB polarization profile is monitored by passing the CVB through a Wollaston polarizer having a $10^{-5}$ extinction ratio, which shows this decomposition (Fig. 1(c)). We show these profiles for vertical and horizontal polarization, and plot the intensity as a function of angular position on the CCD camera. For this paper, the minima are typically <1-2% of the peak intensity. Furthermore, a well-aligned CVB typically has < 3% azimuthal intensity variation, shown in Fig. 1(c).

Once a beam with radial or azimuthal polarization is formed, interconversion is done trivially with bulk waveplates. For example, two half-wave plates oriented by an angle $\theta$ relative to each other rotate an arbitrary input linear polarization by $2\theta$. Hence a radially-polarized beam can be formed by passing a CVB with azimuthal polarization through two half-wave plates with θ=45 degree relative orientation (Fig. 2). Furthermore, a quarter-wave plate can change the CVB from one that contains all linear polarizations (all points on the equator of the Poincaré sphere) to one having all elliptical polarizations (all points on one longitude line of the Poincaré sphere). The polarization profiles and conversions in this work are shown in Fig. 2. Note that, as in Fig. 1(c), because the beam may be considered to be an in-phase superposition of orthogonally-polarized $TEM_{10}$ and $TEM_{01}$ modes, the effect of the quarter-wave plate is to produce a beam that is a superposition of $TEM_{10}$ and $TEM_{01}$ modes with opposite spin.

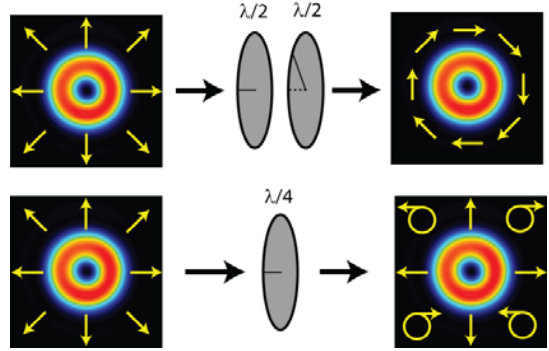

Fig. 2. Polarization profiles used in this work.

## 3. Experiments

Light polarization plays a crucial role in atomic state preparation, manipulation, and detection. Changing the polarization alters the selection rules and transition strengths, but with standard Gaussian beams, which have uniform polarization profile, the effect of polarization can only be studied by adjusting waveplates and other bulk optics components. Spatial dependence of signals occurs through the intensity variation of the Gaussian beam, but this is usually avoided rather than utilized.

With CVBs, polarization effects can be determined rapidly because the sample can be probed with all light polarizations simultaneously. For a uniformly-prepared atomic sample, such as a vapor cell or large ensemble of cold atoms, one can image the effect of incident polarization in a single measurement cycle. This drastically lowers measurement time and improves reproducibility. While it reduces the number of photons available for a given polarization, this is often not the limiting factor in the measurement. One can also prepare a sample with a spatially-varying profile to measure the effect of different spin polarizations quickly. These two uses of CVBs – parallel preparation and detection – are demonstrated in this section.

*3.1 Atomic spin analyzer*

In this subsection, we demonstrate the parallelization of the measurement process by monitoring optical pumping of a $^{85}$Rb vapor cell using a CVB containing all degrees of ellipticity. We observe spatial modulation of the CVB through polarization-dependent absorption, showing that a CVB can serve as a spatial atomic spin analyzer. The experimental setup and beam profiles are shown in Fig. 3. A strong Gaussian pump beam at 795 nm is collimated from a PM optical fiber to a $1/e^2$ radius of 2.6mm with power 1mW (peak intensity 11mW/cm$^2$). This beam passes through a 7.5-cm-long vapor cell, which is heated to 60° C and kept inside a housing formed out of 3-mm-thick μ-metal. A coaxial solenoid surrounding the cell is also inside the housing, and provides a small (<100mG) field as a quantization axis.

The beam is generated by an external cavity diode laser (New Focus Vortex II) and amplified to 100 mW by injection seeding a slave diode. Part of this light passes through the CVB mode converter to produce an azimuthally polarized beam with peak-peak diameter of 1.5mm and power of 1 μW (peak intensity 0.03mW/cm$^2$) at the cell entrance.

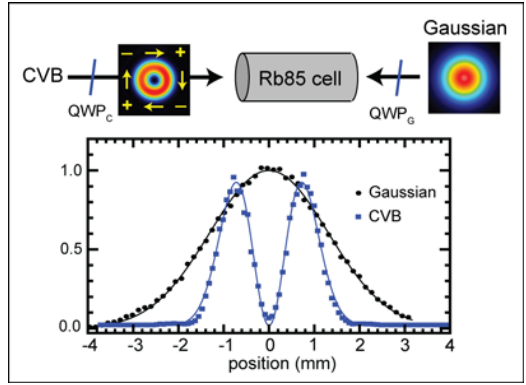

Fig. 3. Top: Setup for optical pumping. The CVB probes the atoms that were opticaly pumped by the Gaussian beam. QWP$_C$: λ/4 plate for the CVB; QWP$_G$: λ/4 plate for the Gaussian. Bottom: Line profiles of the CVB (blue) and Gaussian (black) beams. The $1/e^2$ Gaussian diameter is 5.2 mm; the CVB peak-peak separation is 1.5 mm. Fits to Gaussian and CVB intensity profiles are shown as solid lines.

The Gaussian pump and CVB probe beams are coaxial and counterpropagate, and their frequency is tuned approximately 300 MHz red of the $F$=3 to $F$'=2 transition so that effects of $F$'=3 transitions are reduced. Furthermore, the Gaussian pump contains 3-5% repump light connecting $F$=2 to $F$'=2 by double-passing a 1.5 GHz acousto-optic modulator. To observe selective absorption, we use a Gaussian beam with σ$^-$ polarization. An azimuthally-polarized CVB passes through a rotatable quarter-wave plate (QWP$_C$) prior to entering the cell so that it has position-dependent ellipticity (Fig. 3, top). QWP$_C$ is initially set with its fast axis along the $y$-axis, so that the first and third quadrants contain predominantly σ$^+$ polarization, and the second and fourth quadrants have predominantly σ$^-$ polarization. Rotating this waveplate also rotates this profile: e.g. a 45° clockwise waveplate rotation puts σ$^+$ on the x-axis (to help distinguish angular measurements, we use radians for azimuthal location, and degrees (°) for waveplate orientation). The CVB is imaged onto a CCD camera to observe the absorption.

Figure 4(a) shows the transmitted beam for various orientations of QWP$_C$. For a σ$^-$ Gaussian beam, only the σ$^-$ portions of the CVB, peaking in quadrants 2 and 4, are transmitted. As the waveplate is rotated, the transmitted beam orientation is also rotated. We also show the azimuthal profiles for different orientations to measure the transmission contrast ratio (ratio of maximum to minimum intensity along the azimuthal coordinate), which, for some waveplate orientations, exceeds 50. Some CVB beam orientations are not as pure (see Fig. 1(c)), so the contrast is reduced to 6-7. The transmission contrast for this work is limited by the input beam quality or nonuniform optical pumping. In situations where large magnetic fields exist, however, magneto-optic polarization effects could also become important.

The azimuthal profiles shown in Fig. 4(a) are sinusoidal, indicating that the atomic medium is acting as a fixed, narrowband optical polarizer for the CVB with high extinction ratio. The σ$^-$ components are almost entirely transmitted, and the σ$^+$ components are absorbed by the optical depth of the medium to produce beams similar to a CVB that has passed through a linear polarizer (e.g. Fig. 1(c)). In this configuration it is analogous to a "polarization axis finder" for linearly polarized light.

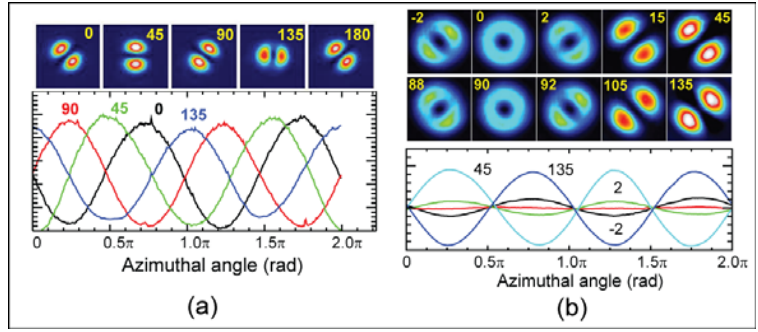

Fig. 4. Optical pumping using a CVB as probe. (a) σ⁻ Gaussian pump beam, probed with a CVB profile similar to that used in Fig. 3. QWP$_C$ is oriented with the fast axis at the orientation shown. Azimuthal intensity profiles for the transmitted CVB probe. (b) Rotation of QWP$_G$ to angles as shown in the images. For 0 degrees, the Gaussian is vertically polarized; at 45 degrees, it has σ⁻ polarization (Media 1).

The orientation of the quarter-wave plate after the Gaussian beam (QWP$_G$) determines the overall degree of spin polarization of the sample, which in turn controls the extinction ratio. Figure 4(b) shows the transmission as a function of the Gaussian beam ellipticity, with QWP$_C$ = 0°. When the Gaussian has σ⁺ (σ⁻) polarization, all atoms are pumped to the $m_F = F$ (-$F$) state, and the σ⁺ (σ⁻) portions of the CVB are transmitted. When the Gaussian pump beam has linear polarization (QWP$_G$=0°, 90°), the σ⁺ and σ⁻ portions are equally transmitted so that no spatial modulation of the beam occurs. Azimuthal profiles of the images are shown. Without analysis, the transmitted CVB images serve as a measure of the net spin analyzer of the ensemble; in principle, a quantitative measure of the spin could be determined through calculations or calibration, but this is outside the scope of this paper.

Using a CVB probe provides common-mode rejection so that comparisons between changes in probe polarization are made rapidly. For example, when the pump polarization is linear, then by symmetry there is no azimuthal intensity variation, as noted in Fig. 4(b). However, deviations from a linear pump are readily detected in a single image as spatial modulation of the probe CVB. As above, this sensitivity depends strongly on the pump power and optical depth of the vapor cell and is outside the scope of this paper, but for our parameters even a 1 degree (17 mrad) rotation of QWP$_G$ resulted in a clear modulation of the output profile. Such relative measurements can be made with standard Gaussian beams and bulk optics, but by putting all polarizations into one beam, the measurement is simultaneous.

### 3.2 Spatially-modulated spin-polarization

If the CVB is significantly stronger than the Gaussian beam, their roles are reversed and the pumping becomes position dependent. This can be used to prepare the atom sample in a spatially varying way. This situation could be particularly useful for multiplexing effects that depend on spin-polarization or nonlinear processes, such as electromagnetically-induced transparency [20, 25], or investigating spatial dynamics in systems with strong atom-atom interactions, such as Bose-Einstein condensates. To demonstrate spatially-modulated spin polarization, we increase the power of the CVB to 1 mW and decrease the power of the Gaussian beam to 10 μW. The Gaussian beam profile at the cell output is imaged onto the CCD camera. The CVB now contains about 5% repump light using a double-passed 1.5 GHz AOM. The cell contains 10 Torr of neon buffer gas to prevent atom movement during the imaging, but to freeze the spatial features further, the Gaussian probe is pulsed for only 2 μs by an AOM. Motion blur becomes noticeable after about 20 μs.

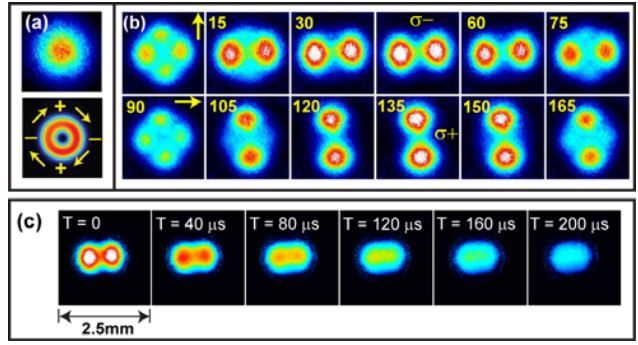

Fig. 5. Top: (a) Top: Gaussian profile; bottom: CVB profile. (b) Gaussian probes with indicated orientation of QWP$_G$. (Media 2) (c) Images of the movement of optically pumped atoms within the buffer gas for 200 μs after the optical pumping pulse is extinguished.

Figure 5(a) shows the Gaussian beam and CVB profiles. QWP$_C$ was rotated so that the locations of pure circular polarization were approximately aligned with the Cartesian axes. The atomic medium, pumped by the CVB, has a spatially-varying spin with $m_F = F$ close to the y-axis, and $m_F = -F$ close to the x-axis. In Fig. 5(b), we show transmitted images (normalized by the Gaussian profile) with different orientations of QWP$_G$. For QWP$_G = 45°$ (135°), the Gaussian probe has $\sigma^-$ ($\sigma^+$) polarization, and the beam is transmitted along the x(y)-axis. In between these circular polarizations, the transmission profile becomes more clover-like as the Gaussian probe beam has equal amounts of $\sigma^+$ and $\sigma^-$. When it is purely linearly polarized (QWP$_G = 0°$, 90°), transmission on these axes are equal. Nodes occur at locations where the CVB is linearly polarized, because at those locations the medium has no net spin polarization, and those portions of the Gaussian are more strongly absorbed.

Because the atoms are contained in a 10 torr buffer gas of neon, these profiles diffuse slowly, and the diffusion can be monitored by pulsing a $\sigma^-$ Gaussian probe beam after shutting off the optical pumping beam. Fig. 5(c) shows images using a 2 μs pulse of the Gaussian beam at different times out to T = 200 μs. The profiles diffuse ~1 mm in 100-200 μs, which is consistent with the 100 μs diffusion time approximated by [26]

$$\tau_{diff} = \frac{a^2}{1.15} \frac{P_{Ne}}{P_{atm}} \qquad (4)$$

where $a$ is the distance in centimeters, $P_{Ne}$ is the neon pressure, $P_{atm}$ is atmospheric pressure, and $\tau_{diff}$ is measured in seconds. Diffusion effects can of course be measured without CVBs, and have been studied in warm vapor [27], but because spatial information is inherent to these experiments, these results are included here.

### 4. Conclusions

We have demonstrated the use cylindrical vector beams for rapid, single-shot atomic spin visualizations in a vapor cell. In uniformly prepared vapors, we have shown that these beams can act as atomic spin analyzers. Additionally, higher-power CVBs were used to spatially modulate the spin of the atoms in the vapor. Temporal dynamics of the spin polarization were recorded by pulsed, uniform-intensity probe beams. We have also discussed the performance of fiber-based CVB generation at rubidium wavelengths. This work was supported by the Office of Naval Research and the Defense Advanced Research Projects Agency. We acknowledge G. Beadie and J. Pechkis for helpful discussions throughout the work.